\documentclass[aip,jcp,graphicx,amsmath,amssymb,reprint,longbibliography]{revtex4-2}

\usepackage{physics}
\usepackage{dsfont}

\usepackage{libertine}
\usepackage[unicode,psdextra]{hyperref}
\usepackage{graphicx}
\usepackage{xcolor}
\hypersetup{
    colorlinks,
    linkcolor={red!50!black},
    citecolor={blue!50!black},
    urlcolor={blue!90!black}
}

\DeclareMathOperator{\prob}{Prob}
\newcommand{\deltadiscr}[1]{\delta\!\left\llbracket #1\right\rrbracket}

\usepackage{enumitem}
\usepackage{stmaryrd}

\newcommand{\unbiaseddistr}{f}
\newcommand{\biaseddistr}{g}
\newcommand{\Obs}{Q}
\newcommand{\obs}{q}

\begin{document}

\title{Importance Sampling for counting statistics in one-dimensional systems}

\author{Ivan N. Burenev} 
\email{inburenev@gmail.com}
\author{Satya N. Majumdar}
\author{Alberto Rosso}
\affiliation{LPTMS, CNRS, Université Paris-Saclay, 91405 Orsay, France}

\date{\today}

\begin{abstract}
	In this paper, we consider the problem of numerical investigation of the counting statistics for a class of one-dimensional systems. Importance sampling, the cornerstone technique usually implemented for such problems, critically hinges on selecting an appropriate biased distribution. 
	While exponential tilt in the observable stands as the conventional choice for various problems, its efficiency in the context of counting statistics may be significantly hindered by the genuine discreteness of the observable.
	To address this challenge, we propose an alternative strategy which we call \emph{importance sampling with the local tilt}.
	We demonstrate the efficiency of the proposed approach through the analysis of three prototypical examples: a set of independent Gaussian random variables, Dyson gas, and Symmetric Simple Exclusion Process (SSEP) with a steplike initial condition. 
\end{abstract}

\maketitle

\section{Introduction}\label{sec:problem}
\par Counting statistics, an important concept in many-body systems, quantifies fluctuations of the number of particles in a given region.
This concept naturally arises in various contexts, spanning both quantum and classical domains. 
In quantum physics, examples include
shot noise \cite{LLL-96,TPPCASM-22,OBS-23}, 
fermionic and spin chains \cite{IA-13,ER-13,SP-17,GEC-18,GLC-20,PPGA-24}, 
trapped fermions \cite{E-13,DDMS-16,LMS-19,DDMG-19,SDMS-21,SDMS-21-SciPost}, 
and quantum dots\cite{GLSSIEDG-06,GMB-06}. 
Counting statistics are relevant in classical physics, for instance, when studying transport problems\cite{BBP-00,M-21} and were thoroughly studied for 
diffusive systems \cite{BD-04,BSGJL-05,DG-09,KM-12,KMS-14,DMS-23},
particles with resetting \cite{BKP-19,DBHMMERS-23,MB-23}, 
active lattice gases \cite{JDR-23,ARKL-23}, 
run and tumble particles \cite{BMRS-20,JRR-23,JRR-23b,BJPR-24},
and exclusion processes \cite{DL-98,DG-09b,LM-11,LMK-24}. 
In addition, counting statistics are applied in closely related problems of one-dimensional gases and random matrix theory \cite{MNSV-09-PRL,NMV-10-PRL,NMV-11,MNSV-11-PRE,MMSV-14,MMSV-16,DKMSS-17-PRL,DKMSS-18-JPA,FMS-22,L-22}.

\par This paper addresses counting statistics for one-dimensional systems while opting for the domain of interest to be a semi-infinite interval. In a rather general form, the problem is formulated as follows. Consider a set of random variables $\mathbf{x} = (x_1,\ldots,x_N)$ with a probability distribution $\unbiaseddistr(\mathbf{x})$. 
Denote by $\Obs(\mathbf{x};z)$, the number of $x_i$ that are greater than some fixed threshold $z$, 
\begin{equation}\label{eq:Q(z)=def}
	\Obs(\mathbf{x};z) = \sum_{i=1}^{N} \theta\left( x_i - z \right), \qquad
	\theta(x) = \begin{cases}
		1, & x\ge 0,\\
		0, & x<0.
	\end{cases} 
\end{equation}
Due to the randomness of $x_i$, quantity $\Obs$ is also a random variable. The goal is to analyze its statistics by studying the probability distribution
\begin{equation}\label{eq:P[q;z]=def}
	\mathrm{P}[\obs;z] \equiv \prob\left[\Obs(\mathbf{x};z)=\obs\right].
\end{equation}
Note that the aim is to explore the full probability distribution rather than focusing on the typical values of $\Obs(\mathbf{x};z)$, where $\mathrm{P}[\obs;z]$ can often be approximated by a normal distribution.

\par 
The technique widely used for capturing rare events in the numerical simulation is \emph{importance sampling} Monte-Carlo\cite{H-02,Bucklew-04,Krauth-06}. This strategy consists of drawing instances of $\mathbf{x}$ from the \emph{biased distribution} that assigns additional statistical weight to rare events. Clearly, the efficiency of this method significantly depends on the choice of the biased distribution.
One conventional choice involves introducing an exponential \emph{tilt in the observable} of interest\cite{EMH-04,H-11,BMRS-20,HDMRS-18,BMRS-19,DBHMMERS-23,BMR-23,BMR-24}. Unfortunately, if this observable is discrete, such as $\Obs(\mathbf{x};z)$, then the \emph{tilt in the observable} may face substantial difficulties (see Sec.\ref{sec:Why discrete is a problem} for the details). 

\par 
Here, we propose an alternative approach that we call \emph{local tilt}. The key idea is to shift the focus from the discrete observable $\Obs(\mathbf{x};z)$ to a continuous (or sufficiently smooth) one. 
To illustrate that this approach is efficient and easy to implement, we examine three systems: a set of independent Gaussian random variables, Dyson gas, and Symmetric Simple Exclusion Process with a steplike initial condition.

\par This paper is organized as follows. In Sec.~\ref{sec:Sampling-scheme} we recall the basics of the importance sampling Monte-Carlo and formulate the \emph{local tilt} strategy.
In Sec~\ref{sec:Why discrete is a problem} we analyze counting for a set of independent Gaussian random variables and explain why the discreteness of $\Obs(\mathbf{x};z)$ may pose a problem. We show how this problem is circumvented by importance sampling with the local tilt.   
Sections \ref{sec:log-gas} and \ref{sec:SSEP} are devoted to the counting statistics of Dyson gas and SSEP, respectively. Finally, we conclude in Sec~\ref{sec:conclusion}.

\section{General importance sampling scheme}\label{sec:Sampling-scheme}
\par Let us first recall the basic scheme that can be implemented to numerically study the probability distribution  $\mathrm{P}[\obs;z]$ of $\Obs(\mathbf{x};z)$. Formally, $\mathrm{P}[\obs;z]$ is obtained from $\unbiaseddistr(\mathbf{x})$ through 
\begin{equation}\label{eq:P[q;z]=int P Q}
	\mathrm{P}[\obs;z] = \int \dd \mathbf{x} \, 
		\unbiaseddistr(\mathbf{x}) \,
		\deltadiscr{\Obs(\mathbf{x};z)-\obs},
\end{equation}
where $\deltadiscr{\obs}$ is a unit sample function
\begin{equation}
	\deltadiscr{\obs}=
	\begin{cases}
		1, & \obs =0,\\
		0, & \obs \ne 0.
	\end{cases}
\end{equation}
Expression \eqref{eq:P[q;z]=int P Q} readily suggests a natural approach. The distribution $\mathrm{P}[\obs;z]$ can be estimated by drawing $n$ instances of $\mathbf{x}$ directly from $\unbiaseddistr(\mathbf{x})$ and then counting how many of them result in the specified value $\obs$, 
\begin{equation}\label{eq:P[q;z]=simple metropolis}
	\mathrm{P}[\obs;z] \approx \frac{1}{n}\sum_{i=1}^n 
		\deltadiscr{\Obs(\mathbf{x}^{(i)};z) - \obs},\quad
	\mathbf{x}^{(i)} \leftarrow \unbiaseddistr(\mathbf{x}) 	
\end{equation}
This strategy, usually referred to as \emph{direct sampling}, is effective close to the typical values of $\Obs(\mathbf{x};z)$.

\par However, the smaller the value of $\mathrm{P}[\obs;z]$, the larger the number of samples required in \eqref{eq:P[q;z]=simple metropolis} for an accurate estimation;
for instance, if the probability of an atypical value of 
$\Obs(\mathbf{x};z)$ is $10^{-10}$, then more than $10^{10}$ samples are necessary. 
This inability to capture rare events makes the direct sampling unsuitable for studying the tails of the distribution, where $\mathrm{P}[\obs;z]$ is usually very small.

\par One of the most effective ways to study the tails of the distribution and capture the rare events is the \emph{importance sampling} Monte Carlo strategy\cite{H-02,Bucklew-04}. The idea is simple.
First, \eqref{eq:P[q;z]=int P Q} can be formally rewritten as
\begin{equation}\label{eq:P[q;z]=int is}
	\mathrm{P}[\obs;z] = \int \dd \mathbf{x}\, 
		\biaseddistr(\mathbf{x})\,
			\frac{\unbiaseddistr(\mathbf{x})}{\biaseddistr(\mathbf{x})}
			\deltadiscr{\Obs(\mathbf{x};z)-\obs},
\end{equation}
where $\biaseddistr(\mathbf{x})$, referred to as \emph{biased distribution}, is an arbitrary probability distribution of $\mathbf{x}$. 
Comparing \eqref{eq:P[q;z]=int P Q} and \eqref{eq:P[q;z]=int is} then suggests that $\mathrm{P}[\obs;z]$ can be estimated by
\begin{equation}\label{eq:P[Q;z]=importance sampling}
	\mathrm{P}[\obs;z] \approx 
		\frac{1}{n}\sum_{i=1}^n 
			\frac{\unbiaseddistr(\mathbf{x}^{(i)})}{\biaseddistr(\mathbf{x}^{(i)})}
			\deltadiscr{\Obs(\mathbf{x}^{(i)};z) - \obs},
		\quad
		\mathbf{x}^{(i)} \leftarrow \biaseddistr(\mathbf{x}). 	
\end{equation}
The difference with respect to \eqref{eq:P[q;z]=simple metropolis} is that the samples of $\mathbf{x}$ are drawn from the biased distribution $\biaseddistr(\mathbf{x})$ and not from the original distribution $\unbiaseddistr(\mathbf{x})$. In addition, the observable is reweighed by $\unbiaseddistr(\mathbf{x})/\biaseddistr(\mathbf{x})$.  

\par Evidently, \eqref{eq:P[q;z]=simple metropolis} and \eqref{eq:P[Q;z]=importance sampling} are equivalent and they both are exact in the limit $n\to\infty$. However, in practice, an appropriate choice of the biased distribution $\biaseddistr(\mathbf{x})$ may tremendously reduce the number of samples required to obtain an accurate estimation of $\mathrm{P}[\obs;z]$. The main challenge is thus to find a suitable biased distribution $\biaseddistr[\mathbf{x}]$.

\par 
Suppose that the goal is to compute the probability that $\Obs(\mathbf{x};z)=\obs_0$; then, the optimal choice for the biased distribution is the conditional distribution,
\begin{equation}\label{eq:Q_best[x;z,q0]}
	\biaseddistr_\text{optimal}(\mathbf{x};z,\obs_0) = \unbiaseddistr\left(\mathbf{x} \,\vert\, \Obs(\mathbf{x};z) = \obs_0\right).
\end{equation}
All instances of $\mathbf{x}$ drawn from \eqref{eq:Q_best[x;z,q0]} result in $\Obs(\mathbf{x};z) = \obs_0$ and hence \eqref{eq:P[Q;z]=importance sampling} is exact for any $n$. 
In practice, determining distribution \eqref{eq:Q_best[x;z,q0]} is akin to the exact analytical computation of $\mathrm{P}[\obs;z]$. 
Since such computations usually are a formidable task, employing \eqref{eq:Q_best[x;z,q0]} becomes unfeasible, thereby necessitating the use of a different biased distribution.

\subsection{Tilt in the observable}

\par A commonly used biased distribution \cite{EMH-04,H-11,BMRS-20,HDMRS-18,BMRS-19,DBHMMERS-23,BMR-23,BMR-24}, which we refer to as \emph{tilt in the observable}, comprises  exponentially biasing the observable of interest
\begin{equation}\label{eq:Q_obs=def}
 	\biaseddistr_\text{obs}(\mathbf{x};\beta) = \frac{1}{Z(\beta)}\,  
 	e^{ \beta \Obs(\mathbf{x};z) } \unbiaseddistr(\mathbf{x}),
 \end{equation} 
where the subscript ``obs'' stands for the ``observable,'' $\beta$ is a parameter and $Z(\beta)$ is the normalization factor,
\begin{equation}
	Z(\beta) = \int \dd \mathbf{x}\, e^{ \beta \Obs(\mathbf{x};z) } \unbiaseddistr(\mathbf{x}).
\end{equation}
A positive (negative) value of $\beta$ in \eqref{eq:Q_obs=def} assigns additional weight to the configurations with atypically large (small) values of $\Obs(\mathbf{x};z)$. 

\par The importance sampling with \eqref{eq:Q_obs=def} is remarkably robust: it may be seamlessly applied to an arbitrary observable and its implementation is quite straightforward, particularly with methods such as the Metropolis-Hastings algorithm \cite{Krauth-06}. However, this strategy may encounter significant challenges when dealing with discrete observables such as $\Obs(\mathbf{x};z)$, as illustrated with a simple example in Sec.~\ref{sec:iidGaussianz=5}.

\subsection{Local tilt}
\par Here, we propose an alternative approach that we call \emph{local tilt}.
It originates from a simple observation. Denote by $M_k[\mathbf{x}]$ the $k$th largest value among the set $(x_1,\ldots,x_N)$. The configurations with $\Obs(\mathbf{x};z)\ge k$ are exactly those with $M_k[\mathbf{x}]\ge z$. Consequently, for $\obs\ge k$,
\begin{multline}\label{eq:P[q,z]=P[q] * P[M]}
	\mathrm{P}[\obs;z] = 
	\prob[M_k[\mathbf{x}] \ge z ]
	\\ \times
	\prob[\Obs(\mathbf{x};z)=\obs \, \vert\, M_k[\mathbf{x}] \ge z] .
\end{multline}
The above equation suggests that studying the distribution $\mathrm{P}[\obs;z]$ with $\obs\ge k$ may be split into two steps as follows:
\begin{itemize}
	\item The first step is to compute the probability $\prob[M_{k}[\mathbf{x}]\ge z]$
	\item The second step is to study the distribution $\prob[\Obs(\mathbf{x};z)=\obs \, \vert\, M_k[\mathbf{x}] \ge k ]$ thereby identifying individual contributions of configurations with different values of $\Obs(\mathbf{x};z)$, i.e., $\Obs(\mathbf{x};z)=k,k+1,$ etc. 
\end{itemize} 
Both steps are conveniently performed by utilizing importance sampling with the biased distribution 
\begin{equation}\label{eq:Q_loc=def}
	\biaseddistr_\text{loc}(\mathbf{x};\gamma,k) = 
	\frac{1}{Z(\gamma, k)}
	e^{ \gamma \min(0, M_k[\mathbf{x}]-z) } 
	\unbiaseddistr(\mathbf{x}),
\end{equation}
where the subscript ``loc'' stands for the ``local,'' $\gamma$ is a parameter, and the normalization factor $Z(\gamma,k)$ is given by
\begin{equation}
	Z(\gamma,k) = \int \dd \mathbf{x} \,
	e^{\gamma \min(0,M_k[\mathbf{x}]-z) } \unbiaseddistr(\mathbf{x}).
\end{equation}

\begin{figure*}[t]
\includegraphics[]{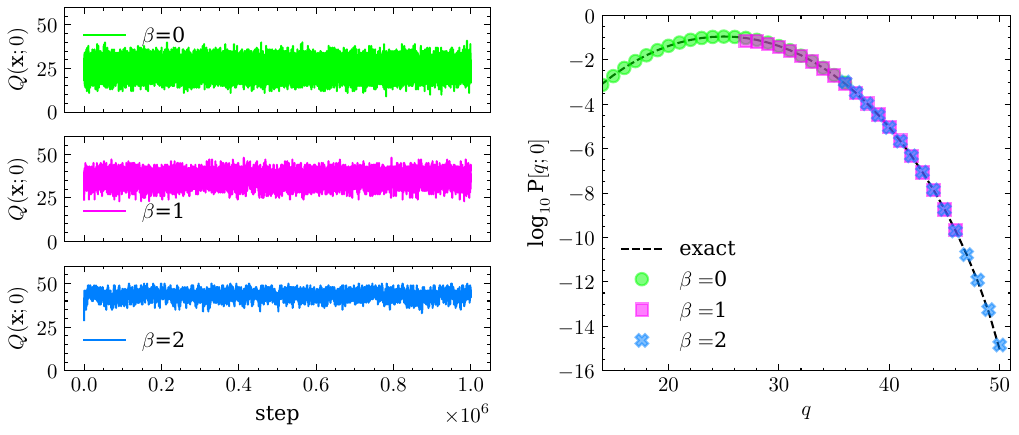}
\caption{Counting statistics for independent Gaussian random variables for $z=0$ and $N=50$. 
Left: Values of $\Obs(\mathbf{x};0)$ obtained using the Metropolis-Hasting algorithm with the tilt in the observable \eqref{eq:Q_obs=def} for various parameters $\beta$. Recall that $\beta=0$ corresponds to an unbiased distribution.
Right: The histogram of $\Obs(\mathbf{x};0)$ computed from the data presented in the left plane. To mitigate the effects of noise, only values of $\Obs(\mathbf{x};0)$ that occurred at least $500$ times are used for the histogram. The dashed line corresponds to the exact result \eqref{eq:app-ex1-binomial}.
}\label{fig:ex1-naive-z=0-Q-data}
\end{figure*}

\par The reasoning behind the choice \eqref{eq:Q_loc=def} is explained in detail in Sec.~\ref{sec:Why discrete is a problem}. The key idea is that it circumvents the issues arising from the discreteness of $\Obs(\mathbf{x};z)$ by focusing on $M_k[\mathbf{x}]$, an observable, which is usually continuous (or ``sufficiently smooth''). The exact form of the bias is owed to the fact that in the limit $\gamma\to\infty$, distribution \eqref{eq:Q_loc=def} becomes the conditional distribution,
\begin{equation}\label{eq:lim Qloc=cond}
	\lim_{\gamma\to\infty} \biaseddistr_\text{loc}(\mathbf{x};\gamma,k) = \unbiaseddistr(\mathbf{x}\,\vert\,\Obs(\mathbf{x};z)\ge k).
\end{equation}
By varying $\gamma$ in \eqref{eq:Q_loc=def} from zero to infinity, we interpolate between the unbiased distribution $\unbiaseddistr(\mathbf{x})$ and the conditional distribution \eqref{eq:lim Qloc=cond}, which is similar to the optimal one \eqref{eq:Q_best[x;z,q0]}.

\par Sections \ref{sec:Why discrete is a problem}-\ref{sec:SSEP} consider three numerical approaches for analyzing the counting statistics: direct sampling, importance sampling with the tilt in the observable, and importance sampling with the local tilt. All three strategies are equivalent and yield exact results as the number of samples tends to infinity; the difference lies in the efficiency. We focus on several prototypical examples and compare the performance of these strategies, demonstrating that importance sampling with local tilt outperforms the other two.

\section{The importance of being discrete, a simple example.}\label{sec:Why discrete is a problem}

\par In this section, we illustrate why the discreteness of $\Obs(\mathbf{x};z)$ may severely hinder importance sampling with the tilt in the observable. To do so, we focus on a simple example and analyze the counting statistics of independent Gaussian random variables. This example is highly instructive, as it allows us to see the issues that arise in relevant physical systems as through the magnifying glass.

\par Let $\mathbf{x}$ be a set of independent Gaussian random variables with zero mean and unit variance, 
\begin{equation}\label{eq:P[x]=iid gaussian}
	\unbiaseddistr(\mathbf{x}) = \frac{1}{(\sqrt{2\pi})^{N}}\exp\left[ - \frac{1}{2} \sum_{i=1}^{N} x_i^2 \right].
\end{equation}
The goal here is to analyze counting statistics of \eqref{eq:P[x]=iid gaussian}, i.e., to compute the probability distribution $\mathrm{P}[\obs;z]$ of $\Obs(\mathbf{x};z)$.  
Due to the independence of  $x_i$, $\mathrm{P}[\obs;z]$ can be computed exactly for the full range of $\obs$ (see Appendix~\ref{sec:app-analytics-1}). This exact result can then be used as a benchmark for the numerical simulations. 

\par We shall focus on $N=50$ and two distinct values of $z$, namely $z=0$ and $z=5$. For $z=0$, the tilt in the observable is an efficient strategy, whereas it turns out to be inadequate for $z=5$. In contrast, importance sampling with the local tilt performs well in both cases.

\subsection{Tilt in the observable}
\par We use the Metropolis-Hastings algorithm to draw samples from $\biaseddistr_\text{obs}(\mathbf{x})$ given by \eqref{eq:Q_obs=def}: at each step, the algorithm proposes the move $\mathbf{x}\mapsto\mathbf{x}'=\mathbf{x} + \Delta\mathbf{x}$ 
and then accepts this move with the probability
\begin{equation}
	p_\text{acc} = \min\left( 1, 
		\frac{\biaseddistr_\text{obs}(\mathbf{x}';\beta)
			 }{\biaseddistr_\text{obs}(\mathbf{x};\beta)}
	\right).
\end{equation} 
Recall that
\begin{equation}
	\frac{\biaseddistr_\text{obs}(\mathbf{x}';\beta)
		 }{\biaseddistr_\text{obs}(\mathbf{x};\beta)} =
	e^{
		-\frac{1}{2} \left( \mathbf{x}'^2- \mathbf{x}^2 \right)
		+ \beta \left( \Obs(\mathbf{x}';z) - \Obs(\mathbf{x};z)\right) }.
\end{equation}
In our simulations, the strategy to propose the move is to draw $\Delta \mathbf{x}$ from the symmetric uniform distribution with a finite support that is chosen in such a way that the acceptance rate is close to $0{.}5$. The system is initialized at $\mathbf{x}=0$.

\subsubsection{Counting statistics for \texorpdfstring{$z=0$}{z=0}}
\par The results for the case $z=0$ are shown in Fig.~\ref{fig:ex1-naive-z=0-Q-data}. The distribution $\mathrm{P}[\obs;0]$ is symmetric, and hence, we focus only on the atypically large values of $\Obs(\mathbf{x};0)$.

\par The Metropolis-Hastings algorithm does not allow us to compute the normalization factor $Z(\beta)$ required to compute the ratio $\unbiaseddistr(\mathbf{x})/\biaseddistr(\mathbf{x})$ in \eqref{eq:P[Q;z]=importance sampling}. However, if the set of parameters $\beta$ in \eqref{eq:Q_obs=def} is chosen appropriately, then the ranges of $\Obs(\mathbf{x};z)$ obtained for consecutive values of $\beta$ overlap.  The ratio $Z(\beta_1)/Z(\beta_2)$ is obtained for two consecutive values $\beta_1$ and $\beta_2$ by matching the histograms in the overlap region. 
Since $Z(\beta=0)=1$, repeating this procedure results in $Z(\beta)$ for all considered values of $\beta$.

\par Fig.~\ref{fig:ex1-naive-z=0-Q-data} clearly indicates that the importance sampling with the tilt in the observable is quite efficient for $z=0$, as it yields $\mathrm{P}[\obs;0]$ across the entire range of $\obs$.

\subsubsection{Counting statistics for \texorpdfstring{$z=5$}{z=5}}\label{sec:iidGaussianz=5}

\begin{figure}[h]
\includegraphics[]{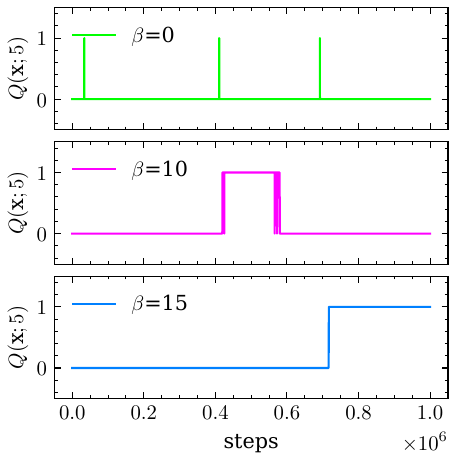}
\caption{Counting statistics for independent Gaussian random variables for $z=5$ and $N=50$. No configuration with $\obs(\mathbf{x};5)>1$ is produced. }\label{fig:ex1-naive-IS-z=5-Q-steps}
\end{figure}

\par From the results shown in Fig.~\ref{fig:ex1-naive-IS-z=5-Q-steps}, it is evident that the case $z=5$ drastically differs from $z=0$: no simulation produces a configuration with $\Obs(\mathbf{x};z)>1$. At the same time, the acceptance rate for all values of $\beta$ is close to $0{.}5$, i.e., the algorithm produces different configurations but all of them results in either $\Obs(\mathbf{x};z)=0$ or $\Obs(\mathbf{x};z)=1$.

\par A possible explanation for such a behavior is that the tilt is not sufficiently strong, and all proposed moves with $\Obs(\mathbf{x};z)>1$ are rejected. However, a more detailed examination reveals that such configurations are never proposed. In particular, this observation implies that changing the exact form of the bias in \eqref{eq:Q_obs=def} from an exponential to any other function of $\Obs(\mathbf{x};z)$ will have no effect.

\begin{figure}[h]
	\includegraphics[width=\linewidth]{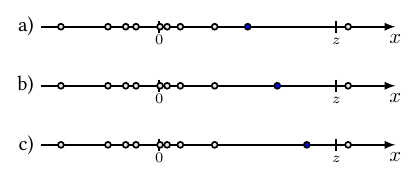} 
	\caption{Three configurations of $\mathbf{x}$ for $N=10$. All configurations correspond to $\Obs(\mathbf{x};z)=1$.  The only difference is in the second largest value $M_2[\mathbf{x}]$ (highlighted in blue).}\label{fig:ex1-configurations-q=0-different}
\end{figure} 

\par 
Three configurations shown in Fig.~\ref{fig:ex1-configurations-q=0-different} help to identify the origin of the problem.
Intuitively, configuration c is somewhat closer to a configuration with the larger $\Obs(\mathbf{x};z)$, in this case $\Obs(\mathbf{x};z)=2$. However, in all three configurations, $\Obs(\mathbf{x};z)=1$; hence, an additional statistical weight corresponding to \eqref{eq:Q_obs=def} remains constant (it is equal to $e^{\beta} / Z(\beta)$).
In other words, the tilt in the observable does not ``push'' the Metropolis-Hastings algorithm towards the configuration with $\Obs(\mathbf{x};z)=2$, and importance sampling with the tilt in the observable is essentially the same as the direct sampling. This implies that if two consecutive values of $\Obs(\mathbf{x};z)$ have very different probabilities, say $10^{-5}$ and $10^{-10}$, then more than $10^{5}$ samples are required to propose the configuration with a larger value of $\Obs(\mathbf{x};z)$.

\par It is reasonable to hope that the issue described above might be resolved by utilizing the biased distribution that distinguishes between configurations in Fig.~\ref{fig:ex1-configurations-q=0-different}. 
The straightforward approach is to assign an additional weight $e^{\gamma M_2[\mathbf{x}]}$ to each configuration. 
Such a biased distribution would clearly steer the algorithm towards configurations where $\Obs(\mathbf{x};z)=2$. 
However, when two configurations with $M_2[\mathbf{x}]\ge z$ are produced, there is no apparent reason to prefer one over the other. Consequently, it is more appropriate to assign a weight $e^{\gamma\min(0,M_2[\mathbf{x}]-z)}$ instead, i.e., to use the local tilt defined in~\eqref{eq:Q_loc=def}.

\begin{figure*}[t]
	\includegraphics[]{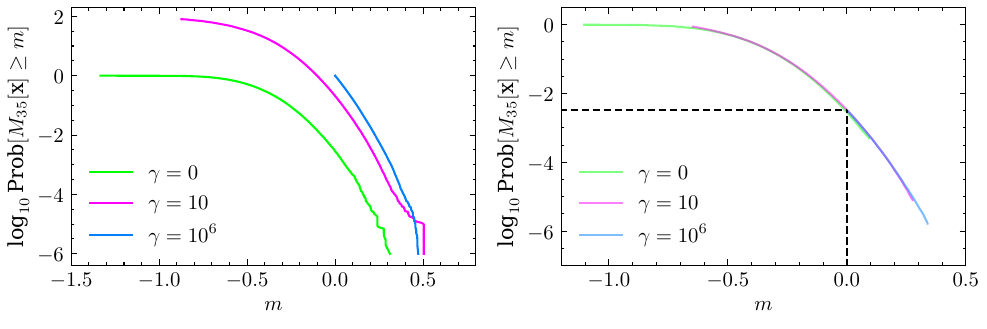}
	\caption{
	Counting statistics for independent Gaussian random variables for $z=0$ and $N=50$ with the local tilt. Computing the first factor in \eqref{eq:P[q,z]=P[q>35] * P[M_35>0] ex1}. The plots of $\prob[M_{35}[\mathbf{x}]\ge m]$ estimated as in \eqref{eq:P[M[x]>ml] = cumsum}. Left: Raw data without reweighing by $Z(\gamma,35)$. Right: Data with normalization factors $Z(\gamma,35)$ restored by matching the different parts of the plot. To reduce the impact of the noise, when matching the plots, the first and last $500$ points are removed. }
	\label{fig:ex1-local-IS-z=0-k=35}
\end{figure*}

\subsection{Local tilt}

\par Now we implement the importance sampling with the local tilt \eqref{eq:Q_loc=def}. The configurations of $\mathbf{x}$ are drawn from $\biaseddistr_\text{loc}(\mathbf{x};\gamma,k)$ by using the Metropolis-Hastings algorithm. The move $\mathbf{x}\mapsto\mathbf{x}' = \mathbf{x}+\Delta\mathbf{x}$ is accepted with probability 
\begin{equation}
	p_\text{acc} = \min\left( 1, 
		\frac{\biaseddistr_\text{loc}(\mathbf{x}';\gamma,k)
			 }{\biaseddistr_\text{loc}(\mathbf{x};\gamma,k)}
	\right).
\end{equation}
Recall that
\begin{multline}
 	\frac{\biaseddistr_\text{loc}(\mathbf{x}';\gamma,k)
			 }{\biaseddistr_\text{loc}(\mathbf{x};\gamma,k)}
	= 
	e^{
		-\frac{1}{2} \left( \mathbf{x}'^2- \mathbf{x}^2 \right) }
	\\
	\times e^{
		\gamma \left( \min(0, M_k[\mathbf{x}']-z) 
					   - \min(0, M_k[\mathbf{x}]-z)
			 \right)}.
\end{multline} 
The strategy to propose the move is again to draw $\Delta \mathbf{x}$ from the symmetric uniform distribution.

\subsubsection{Counting statistics for \texorpdfstring{$z=0$}{z=0}}

\begin{figure}[h]
	\includegraphics{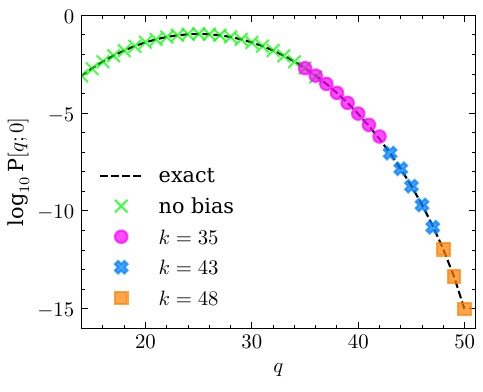}
	\caption{Counting statistics for independent Gaussian random variables for $z=0$ and $N=50$ with the local tilt. The distribution $\mathrm{P}[\obs;z]$. Different parts of the distribution are obtained from $\biaseddistr_\text{loc}(\mathbf{x};\gamma,k)$ for different values of $k$. The dashed line corresponds to the exact result \eqref{eq:app-ex1-binomial}. }
	\label{fig:ex1-local-IS-z=0-Q-hist}
\end{figure}

\par Since direct sampling (see Fig.~\ref{fig:ex1-naive-z=0-Q-data}) suggests that atypically large values are roughly speaking $\Obs(\mathbf{x};0)\ge 35$, we start with studying the distribution $\mathrm{P}[\obs;0]$ for $\obs \ge 35$. Recall that according to \eqref{eq:P[q,z]=P[q] * P[M]}, this probability is represented as a product of two factors. Specifically,
\begin{multline}\label{eq:P[q,z]=P[q>35] * P[M_35>0] ex1}
 	\mathrm{P}[\obs;0] = 
	\prob[M_{35}[\mathbf{x}] \ge 0]
	\\ \times
	\prob[\Obs(\mathbf{x};0)=\obs \, \vert\, M_{35}[\mathbf{x}] \ge 0 ].
\end{multline}

\par The procedure commences with sampling configurations from the biased distribution $\biaseddistr_\text{loc}(\mathbf{x};\gamma,k)$ with a fixed value of $k$; here we set $k=35$. The parameter $\gamma$ is selected from a set of values ranging from zero to a very large number, say $10^{6}$, and an independent simulation is performed for each value of $\gamma$. In the present case, $\gamma\in\{0,10,10^{6}\}$ and $n=10^6$ instances of $\mathbf{x}$ are drawn.  
These data are then used to compute both factors in \eqref{eq:P[q,z]=P[q>35] * P[M_35>0] ex1}.

\par The first factor, $\prob[M_{35}[\mathbf{x}]\ge 0]$, is computed by studying the complimentary cumulative distribution function as
\begin{equation}\label{eq:P[M[x]>m]=def}
	\prob[M_{35}[\mathbf{x}]\ge m] =
	\int \dd \mathbf{x} \, f(\mathbf{x}) \, \theta(M_{35}[\mathbf{x}] - m),
\end{equation}
which can be estimated as in \eqref{eq:P[Q;z]=importance sampling} with $\deltadiscr{Q(\mathbf{x}^{(i)};z)-q}$ replaced by $\theta(M_{35}[\mathbf{x}^{(i)}]-m)$. 
In practice, however, it is more convenient to first sort the values $m_j$ of $M_{35}[\mathbf{x}^{(i)}]$ in the decreasing order $m_1\ge m_2\ge \cdots m_n$, and then estimate the cumulative distribution function as
\begin{equation}\label{eq:P[M[x]>ml] = cumsum}
	\prob[M_{35}[\mathbf{x}]\ge m_\ell] \approx 
	 \frac{Z(\gamma,35)}{n}
	\sum_{i=1}^{\ell} e^{-\gamma \min(m_i-z,0) } .   	
\end{equation}

Using \eqref{eq:P[M[x]>ml] = cumsum} for different values of $\gamma$ yields different parts of $\prob[M_{35}[\mathbf{x}]\ge m]$ (Fig.~\ref{fig:ex1-local-IS-z=0-k=35} left). The factors $Z(\gamma,35)$ are again determined by matching the plots of $\prob[M_{35}[\mathbf{x}]\ge m]$ (Fig.~\ref{fig:ex1-local-IS-z=0-k=35} right). We find that $\prob[M_{35}[\mathbf{x}]\ge 0] \approx 3{.}3\times 10^{-3}$.

\par To compute the second factor in \eqref{eq:P[q,z]=P[q>35] * P[M_35>0] ex1}, $\prob[\Obs(\mathbf{x};z)=\obs\,\vert\, M_{35}[\mathbf{x}]\ge0]$, all the configurations with $M_{35}[\mathbf{x}]\ge 0$ are used without any reweighing to make a histogram  (see Appendix~\ref{sec:app-additional-plots}, Fig.~\ref{fig:ex1-local-IS-z=0-k=35-Q-conf}). Recall that the local tilt \eqref{eq:Q_loc=def} does not impact the relative probabilities of configurations with $M_{35}[\mathbf{x}]\ge 0$.

\par Finally, the distribution $\mathrm{P}[\obs;z]$ is obtained according to \eqref{eq:P[q,z]=P[q>35] * P[M_35>0] ex1} by multiplying the histogram by $\prob[M_{35}[\mathbf{x}]\ge 0]$.
Importance sampling with the local tilt for $k=35$ results in the probability distribution $\mathrm{P}[\obs;z]$ up to $\Obs(\mathbf{x};0)=42$ (see Fig.~\ref{fig:ex1-local-IS-z=0-k=35-Q-conf}). To explore the tail of $\mathrm{P}[\obs;0]$ further, we repeat the above procedure for different values of $k$, eventually obtaining $\mathrm{P}[\obs;0]$ for the entire range of $\obs$ (see Fig.~\ref{fig:ex1-local-IS-z=0-Q-hist}).

\par For the counting statistics of $\Obs(\mathbf{x};z=0)$, importance sampling with the tilt in the observable and importance sampling with the local tilt are viable. The performances are similar, as both approaches require producing approximately $10^6$ to $10^{7}$ configurations.

\subsubsection{Counting statistics for \texorpdfstring{$z=5$}{z=5}}
\begin{figure}[h]
	\includegraphics{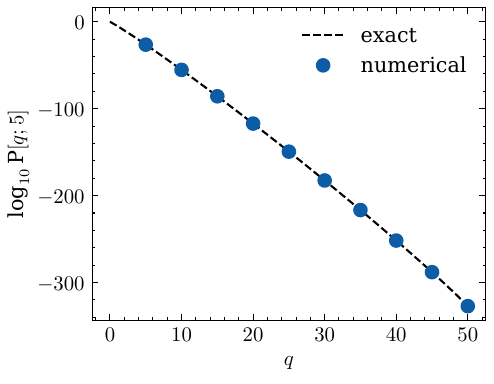}
	\caption{Counting statistics for independent Gaussian random variables for $z=5$ and $N=50$ with the local tilt. The distribution $\mathrm{P}[\obs;5]$. Each point is obtained from an independent set of simulations with $k=\obs$. The dashed line corresponds to the exact result \eqref{eq:app-ex1-binomial}. }
	\label{fig:ex1-local-IS-z=5-Q-hist}
\end{figure}

\par To illustrate that the local tilt is more efficient for $z=5$, we concentrate on $\Obs(\mathbf{x};z)\ge5$. Recall that these configurations were unreachable when using importance sampling with the tilt in the observable. 

\par The procedure is the same as in the case $z=0$. First, $n=10^6$ instances of $\mathbf{x}$ are drawn from $\biaseddistr_\text{loc}(\mathbf{x};\gamma,k=5)$ with $\gamma \in \{ 0,10,20,10^{6} \}$. The data are used to compute $\prob[M_{5}[\mathbf{x}]\ge5]$ as in \eqref{eq:P[M[x]>ml] = cumsum}; the resulting plots are shown in Appendix~\ref{sec:app-additional-plots} (Fig.~\ref{fig:ex1-local-IS-z=5-k=5}).  We find that $\prob[M_5[\mathbf{x}]\ge 5]\approx 4\times10^{-27}$. 
Then, as prescribed, only configurations with $\Obs(\mathbf{x};5)\ge 5$ are left. However, in the case $z=5$, the algorithm does neither produce nor  propose configurations with $\Obs(\mathbf{x};5)\ge 6$. This implies that
\begin{equation}\label{eq:prob[q>5]=prob[q=5]}
	\mathrm{P}[\obs,5] \approx \prob[\Obs(\mathbf{x};5)\ge \obs],
\end{equation}
and that the importance sampling with the local tilt produces a single point $\obs=k$. The histogram of $\mathrm{P}[\obs;z]$ is then obtained by repeating this procedure for different values of $k$ (see Fig.~\ref{fig:ex1-local-IS-z=5-Q-hist}).

\par The importance sampling with the local tilt, allows us to explore $\mathrm{P}[\obs;5]$ for the entire range of $\obs$, and each point on the histogram requires sampling approximately $10^{7}$ instances of $\mathbf{x}$. Thus, the importance sampling with the local tilt clearly outperforms the importance sampling with the tilt in the observable.

\section{Correlated identically distributed random variables: Dyson gas.}\label{sec:log-gas}

\begin{figure}[h]
	\includegraphics{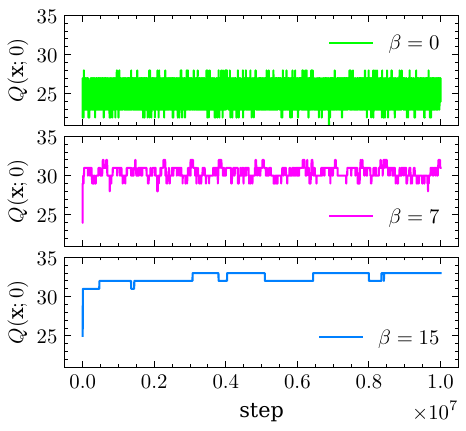}
	\caption{Counting statistics for the Dyson gas with $N=50$ particles for $z=0$. 
	The values of $\Obs(\mathbf{x};0)$ are obtained using the Metropolis-Hastings algorithm with the tilt in the observable \eqref{eq:Q_obs=def} for various parameters $\beta$. }
	\label{fig:ex2-Q-steps}
\end{figure}

\par In this section, we focus on the counting statistics for an $N\times N$ random matrix belonging to the Gaussian orthogonal ensemble. The eigenvalues of such matrices can be represented as a \emph{Dyson gas}, i.e., a system of particles on the line with logarithmic interaction \cite{Forrester-10,Mehta-04}. The joint probability distribution of the eigenvalues $x_i$ reads 
\begin{equation}\label{eq:P[x]=dyson log gas}
	\unbiaseddistr(\mathbf{x}) = \frac{1}{Z_N} 
	\exp\left[ - \sum_i \frac{x_i^2}{2} + \sum_{j<k}\log\abs{x_j-x_k} \right].
\end{equation}
The normalization constant $Z_N$ is known \cite{Mehta-04,Forrester-10}, and for large $N$ it behaves\cite{DM-06-PRL,DM-08-PRE} as $Z_N\sim\exp[ - \Omega_0 N^2]$, where $\Omega_0 = (3 + 2\log 2)/8$.  

\par We focus on the counting statistics for the system of $N=50$ particles for $z=0$. 
To benchmark numerical simulations, we compare them with the analytical expressions\cite{MNSV-09-PRL,MNSV-11-PRE} for the asymptotic behavior of $\mathrm{P}[\obs;0]$ as $N\to\infty$ (see Appendix~\ref{sec:app-analytics-2}).
The configurations are again generated with the same Metropolis-Hastings algorithm as the one in the Sec.~\ref{sec:Why discrete is a problem}. The only difference is that when proposing a move  $\mathbf{x}\mapsto\mathbf{x}'$, only $10$ out of $50$ values of $x_i$ are modified.

\begin{figure}[h]
	\includegraphics{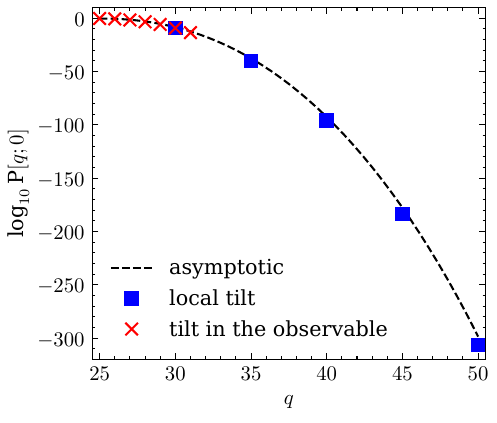}
	\caption{Counting statistics for the Dyson gas with $N=50$ particles for $z=0$. The distribution $\mathrm{P}[\obs;0]$ is obtained using the importance sampling with the tilt in the observable (crosses) and with the local tilt (squares).
	The black dashed line corresponds to the asymptotic behavior of $\mathrm{P}[\obs;0]$ as in \eqref{eq:app-ex2-LDF}.
	Tilt in the observable: we used $\beta \in \{0,3,5,7,10,15\}$ and $10^7$ of $\mathbf{x}$ instances are drawn for each $\beta$. 
	Local tilt: each point is obtained from an independent set of  simulations with $k=\obs$; approximately $10^{7}$ instances of $\mathbf{x}$ are used for each point. }
	\label{fig:ex2-Q-hist}
\end{figure}

\par The results of the numerical simulation with importance sampling with the tilt in the observable \eqref{eq:Q_obs=def} are shown in Fig.~\ref{fig:ex2-Q-steps}. For $\beta=15$, the algorithm produces configurations with $\Obs(\mathbf{x};0) \in \{32,33\}$. Recall that an important step in obtaining the probability distribution $\mathrm{P}[\obs;0]$ involves matching the histograms obtained from different values of $\beta$ in the region of overlap, which now has to be done with only two points. 
Such a behavior suggests that the importance sampling with the tilt in the observable allows us to trustfully explore $\mathrm{P}[\obs;0]$ up to $\obs=33$, unless more samples are used.

\par However, importance sampling with the local tilt allows us to explore the distribution $\mathrm{P}[\obs;0]$ for the entire range of $\obs$ (see Fig.~\ref{fig:ex2-Q-hist}).

\section{Trajectories of the non-crossing particles: Symmetric Simple Exclusion process}\label{sec:SSEP}
\par In previously considered cases, the efficiency of the local tilt is owed to the shifting of the focus from the discrete observable $\Obs(\mathbf{x};z)$ to a continuous $M_k[\mathbf{x}]$. 
However, after a suitable modification, the method may be applied to the systems where $x_i$, and hence $M_k[\mathbf{x}]$, are also genuinely discrete. Such systems appear, for example, in the context of random walks on the lattice\cite{J-02}, growth processes\cite{HZ-95,J-00,PS-00,S-06},  vertex models\cite{Z-00,CP-13,BCMP-23} or exclusion processes\cite{D-07,M-15,KK-10}. As a prototypical example, we consider the Symmetric Simple Exclusion Process (SSEP) with a steplike initial condition\cite{DG-09,DG-09b}.

\par
\begin{figure}[h]
	\includegraphics{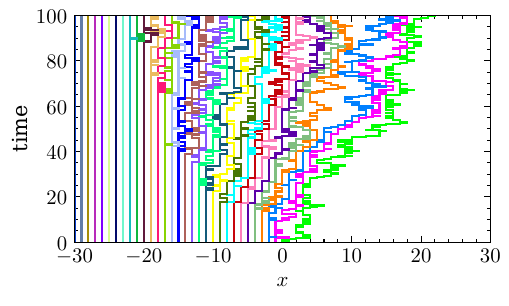}
	\caption{A typical configuration of SSEP with steplike initial conditions. The trajectories of the first 30 particles are shown.}
	\label{fig:ex3-SSEP-typical-conf}
\end{figure}

 Consider a system of particles on the infinite one-dimensional lattice with an exclusion rule that prevents two particles from occupying the same site.
Initially, all sites to the left of the origin (inclusive) are occupied and all sites to the right of the origin are empty. The system then evolves for a time $t$ according to the following stochastic dynamics: each particle has two exponential clocks with rate $1$, one for moving left and the other for moving right. 
When a clock ticks, the particle hops to the adjacent site in the corresponding direction, provided that the target site is empty. 
An example of a configuration is shown in Fig.~\ref{fig:ex3-SSEP-typical-conf}.

\par The goal is to study the statistics of the current, i.e., the total flux of particles between site $0$ and site $1$ up to time $t$, here $t=100$. 
As was noted in Ref.~[\onlinecite{BMRS-20}], for a steplike initial condition, the current is equal to the number of particles located to the right of the origin \emph{at} time $t$. Therefore, the total flux is just $\Obs(\mathbf{x};1)$, where $\mathbf{x}$ denotes the coordinates of the particles at time $t$.
The distribution of $\Obs(\mathbf{x};1)$ in the limit $t\to\infty$ was computed in Ref.~\onlinecite{DG-09b}; here this result is used as a benchmark for the numerical simulations (see Appendix~\ref{sec:app-analytics-3}).

\par Contrary to the previously considered cases, there is no explicit expression for $\unbiaseddistr(\mathbf{x})$. Consequently, obtaining an instance of $\mathbf{x}$ necessitates generating the full evolution of the system. 
To model an infinite system, $N=100$ particles are considered, with the constraint of prohibiting the leftmost particle to hop leftward, i.e., it has only the clock corresponding to the rightward jumps. 
The evolution is simulated using the Metropolis algorithm.
To propose a move, the algorithm selects $r$ out of $N$ particles, then chooses the hopping direction for each particle and resamples $p$ of the time intervals of the corresponding clock. 
The particles, hopping directions, and time intervals are chosen randomly. 
If the sum of the new intervals exceeds the observation time $t$, the sequence is truncated. Conversely, if the sum is smaller, additional intervals are generated. The parameters $r$ and $p$ are tuned to achieve an acceptance rate close to $50\%$; typically they both range from $5$ to $20$. 
Finally, in each simulation, the leftmost particle is checked to ensure that it has not hopped to the right, thereby justifying that the system of $100$ particles is effectively infinite.

\par
\begin{figure}[h]
	\includegraphics{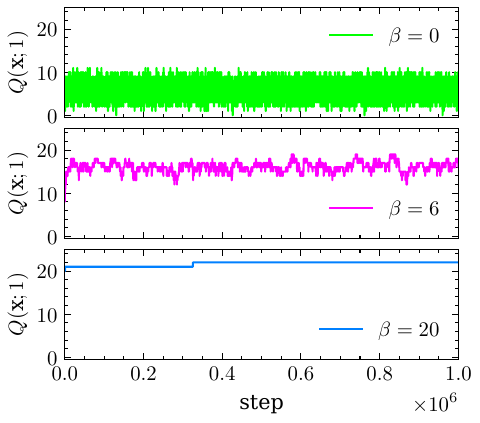}
	\caption{Counting statistics for SSEP with steplike initial conditions at $t=100$. The values of $\Obs(\mathbf{x};1)$ are obtained using the Metropolis-Hastings algorithm with the tilt in the observable \eqref{eq:Q_obs=def} for various parameters $\beta$.}
	\label{fig:ex3-SSEP-Q-data}
\end{figure}

 Importance sampling with the tilt in the observable is implemented in the exact same manner as in the previous examples. 
The results shown in Fig.~\ref{fig:ex3-SSEP-Q-data}	suggest that the tilt in the observable may be used to explore the probability distribution up to $\Obs(\mathbf{x};1)=19$. 

\par From Fig.~\ref{fig:ex3-SSEP-typical-conf}, it is clear that if the particle is deep into the bulk, then due to the exclusion rule, it does not move. Therefore, the step of the local tilt strategy that requires studying the distribution of $M_k[\mathbf{x}]$ should be slightly modified.

\par The trajectories are non-crossing; hence, $M_k[\mathbf{x}]=x_k$ and $\prob[\Obs(\mathbf{x};1) \ge k] = \prob[x_k \ge 1]$ (particles are numbered right to left). The exclusion rule implies 
\begin{equation}\label{eq:prob[Q]=prob[M,M,M,M]}
	\prob[x_k \ge 1] \\
	= \prob\left[
		x_k \ge 1, x_{k-1}\ge 2, \ldots, x_1 \ge k
	\right].
\end{equation}
The above equation suggests that $\prob[x_k\ge 1]$ can be represented as a product of several probabilities. For instance, if $k=30$, then we use
\begin{equation}\label{eq:P[x_30]=split}
\begin{aligned}
 	\prob[x_{30}\ge 1] = & \prob[x_{30}\ge 1\,\vert\, x_{20}\ge 11]
 		\\
 		& \times \prob[x_{20}\ge 11 \,\vert\, x_{10}\ge 21]
 		\\
 		& \times \prob[x_{10}\ge 21].
\end{aligned}
\end{equation}
Each factor in \eqref{eq:P[x_30]=split} is computed separately with the local tilt strategy: first, by biasing $x_{10}$, we compute $\prob[x_{10}\ge 21]$, and then, the conditional probabilities $\prob[x_{20}\ge 11 \,\vert\, x_{10}\ge 21]$ and $\prob[x_{30}\ge 1\,\vert\, x_{20}\ge 11]$ are computed successively by utilizing the local tilt for $x_{20}$ and $x_{30}$.

\begin{figure}[h]
	\includegraphics{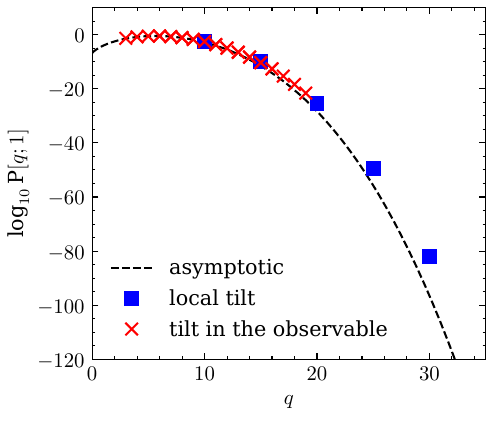}
	\caption{ Counting statistics for SSEP with a steplike initial condition at $t=100$. 
	The distribution $\mathrm{P}[\obs;0]$ is obtained using the importance sampling with the tilt in the observable (crosses) and with the local tilt (squares).
	The black dashed line corresponds to the asymptotic behavior of $\mathrm{P}[\obs;1]$ in the limit $t\to\infty$ as in \eqref{eq:app3-ex3-P[q,1]=}.
	Tilt in the observable: the values $\beta \in \{0,2,4,6\}$ are used and $10^6$ instances of $\mathbf{x}$ are drawn for each $\beta$. 
	Local tilt: each point is obtained from an independent set of  simulations with $k=\obs$; approximately $10^{6}$ instances of $\mathbf{x}$ are used for each point. }
	\label{fig:ex3-Q-hist}
\end{figure}

The results of the simulation shown in Fig.~\ref{fig:ex3-Q-hist} demonstrate that the local tilt again outperforms the tilt in the observable. A configuration with atypically large current is shown in Fig.~\ref{fig:ex3-SSEP-conf-atyp}.


\section{Conclusion}\label{sec:conclusion}

\par 
We have proposed a strategy to numerically study the counting statistics of one-dimensional systems which we call \emph{local tilt}. Our idea involves utilizing importance sampling with the biased distribution that clearly distinguishes between any two configurations typically proposed by a Metropolis algorithm. 

\par  
We analyzed three prototypical systems: a set of independent Gaussian random variables, Dyson gas, and SSEP with a steplike initial condition.  
We have demonstrated that in all three systems, when compared to the importance sampling with the tilt in the observable, the proposed \emph{local tilt} approach maintains the simplicity of the implementation while significantly outperforming the standard technique.

\par There are many questions that remain to be explored further.
The local tilt approach is highly versatile and can be straightforwardly applied to any one-dimensional system where $x_i$ are either identically distributed or genuinely ordered. For instance, it can be used to study the counting statistics of gases with arbitrary pairwise interactions or current fluctuations in systems of stochastic particles with arbitrary dynamics, provided that either the initial conditions are symmetric or a non-crossing constraint is imposed (single-file diffusion). It would be interesting to evaluate the performance of the local tilt approach in these contexts.

\par 
Further investigation is warranted to determine if the local tilt approach can be modified to analyze other observables, such as the number of $x_i$'s in an arbitrary segment. 
Moreover, issues similar to those addressed in this paper may arise even for continuous observables. 
Consider the SSEP discussed in Section~\ref{sec:SSEP}, where a steplike initial condition evolves over time~$t$. Suppose that the goal is to study the statistics of the time spent by the particles in the region $x > 100$. Although this observable is continuous, it typically equals zero for most configurations, as is evident from Fig.~\ref{fig:ex3-SSEP-typical-conf}. This makes it highly unlikely that importance sampling with the tilt in the observable will successfully yield the desired probability distribution. Therefore, it would be intriguing to explore whether the local tilt approach can be modified to effectively handle such problems.


\begin{appendix}

\section{Additional figures}\label{sec:app-additional-plots}

\par In this appendix, we provide several additional figures.
\par
\begin{figure}[ht]
	\includegraphics{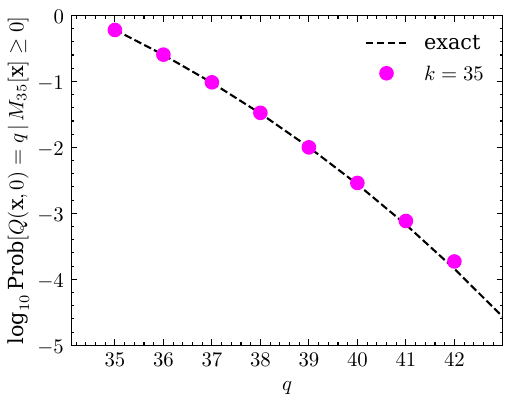}
	\caption{
	Counting statistics for independent Gaussian random variables for $z=0$ and $N=50$ with the local tilt. Computing the second factor in \eqref{eq:P[q,z]=P[q>35] * P[M_35>0] ex1}, 
	the conditional distribution $\prob[\Obs(\mathbf{x};0=q\,\vert\, M_{35}[\mathbf{x}] \ge 0]$.
	The simulations performed with $\gamma=0, 10$ and $10^{6}$ produce $3168$, $206\,239$ and $999\,974$ configurations with $M_{35}[\mathbf{x}]\ge 0$, respectively. These configurations ($1\,209\,381$ in total) are then used without reweighing to construct the histogram. Dashed line corresponds to the exact result \eqref{eq:app-ex1-binomial}.
	}
	\label{fig:ex1-local-IS-z=0-k=35-Q-conf}
\end{figure}

\begin{figure}[ht]
\includegraphics{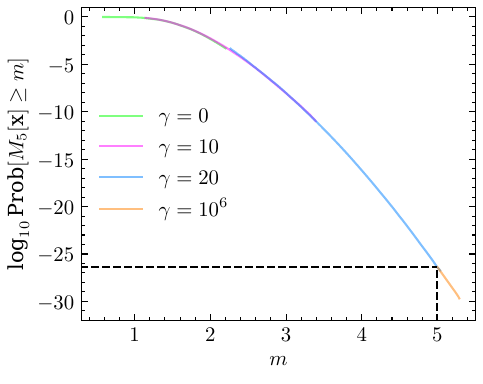}
\caption{Counting statistics for independent Gaussian random variables for $z=5$ and $N=50$ with the local tilt. Computing the first factor in \eqref{eq:P[q,z]=P[q>35] * P[M_35>0] ex1}, $\prob[M_{5}[\mathbf{x}]\ge5]$, estimated as in \eqref{eq:P[M[x]>ml] = cumsum}. 
The normalization factors $Z(\gamma,5)$ restored by matching the different parts of the plot.
To reduce the impact of the noise, when matching the plots, first and last $500$ points are removed. }
\label{fig:ex1-local-IS-z=5-k=5}
\end{figure}

 For the counting statistics of $N = 50$ independent Gaussian random variables with a local tilt, we present the conditional probability distribution $\prob[\Obs(\mathbf{x};0=q\,\vert\, M_{35}[\mathbf{x}] \ge 0]$ (Fig.~\ref{fig:ex1-local-IS-z=0-k=35-Q-conf}) and the complementary cumulative probability distribution of $M_5[x]$ (Fig.~\ref{fig:ex1-local-IS-z=5-k=5}).
\par
\begin{figure}[ht]
	\includegraphics{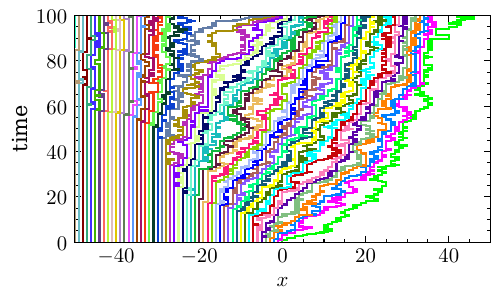}
	\caption{A configuration of SSEP with steplike initial conditions corresponding to the atypical value $Q(\mathbf{x};1)=30$. The trajectories of the first 50 particles are shown.}
	\label{fig:ex3-SSEP-conf-atyp}
\end{figure}

 
For the SSEP with a steplike initial condition, we provide an example of configuration corresponding to the atypical value $Q(\mathbf{x};1)=30$ (Fig.~\ref{fig:ex3-SSEP-conf-atyp}).

\clearpage

\section{Analytical expressions}\label{sec:app-analytics}
\par In this appendix, we give analytical expressions used through the paper. 
\subsection{Example 1. Independent Gaussian random variables}\label{sec:app-analytics-1}
\par For the system of independent Gaussian random variables, the probability distribution $\mathrm{P}[\obs;z]$ is a binomial distribution
\begin{equation}\label{eq:app-ex1-binomial}
	\mathrm{P}[\obs;z] = \frac{N!}{\obs!(N-\obs)!}\, s^\obs (1-s)^{N-\obs} ,
\end{equation}
where $s$ is
\begin{equation}\label{eq:app-ex1-binomial-s=}
	s = \frac{1}{\sqrt{2\pi}} 
		\int_{z}^{\infty} e^{-\frac{x^2}{2}}\dd x.
\end{equation}

\subsection{Example 2. Dyson gas}\label{sec:app-analytics-2}
\par For the Dyson gas \eqref{eq:P[x]=dyson log gas}, at large $N$ the probability distribution $\mathrm{P}[\obs;0]$ admits the large deviation form\cite{MNSV-09-PRL,MNSV-11-PRE}, specifically
\begin{equation}\label{eq:app-ex2-LDF}
 	\mathrm{P}[\obs;0] \simeq \exp\left[ - N^2 \; \Phi\left(\frac{\obs}{N}\right) \right],
 	\quad N\to\infty.
\end{equation}
In terms of the scaling variable 
\begin{equation}
	c = \frac{\obs}{N}, 
\end{equation}
the rate function $\Phi(c)$ is given by
\begin{multline}\label{eq:app-ex2-Phi(c)=}
	\Phi(c) = 
	\frac{1}{4}\left[	L^2 - 1 - \log(2L^2) \right] 
	\\
	+ \frac{1-c}{2} \log a
	- \frac{(1-c) (a^2 - 1)}{4a^2} L^2
	\\
	+ \frac{c}{2} \int_{L}^{\infty} W_1(x)\, \dd x
	+ \frac{1-c}{2} \int_{L/a}^{\infty} W_2(x)\,\dd x,
\end{multline}
where the parameter $a$ is determined implicitly as a function of $c$ as
\begin{equation}\label{eq:app-ex2-a-c}
	\int_{0}^{1} \dd y \sqrt{\frac{1-y}{y}} \sqrt{y^2 +y + \frac{a-1}{a^2}}
	= \frac{\pi}{2} \left(1- \frac{a-1}{a^2}\right) c.
\end{equation}
The parameter $L$ is 
\begin{equation}\label{eq:app-ex2-L(a)=}
	L \equiv L(a) = \frac{a\sqrt{2}}{\sqrt{a^2 - a + 1}}
\end{equation}
and $W_{1,2}(x)$ are
\begin{align}\label{eq:app-ex2-W1=}
	W_1(x) & = x - \frac{1}{x} 
		- \sqrt{\frac{x-L}{x}
				\left(x+\frac{L}{a}\right)
				\left(x + \left(1-\frac{1}{a}\right)L\right) },\\
	\label{eq:app-ex2-W2=}
	W_2(x) & = x - \frac{1}{x} 
		- \sqrt{\frac{x+L}{x}
				\left(x-\frac{L}{a}\right)
				\left(x - \left(1-\frac{1}{a}\right)L\right) }.	
\end{align}

\subsection{Example 3. SSEP}\label{sec:app-analytics-3}
For the Symmetric Simple Exclusion Process as $t\to\infty$, the probability distribution $\mathrm{P}[\obs;1]$ has the form\cite{DG-09b}
\begin{equation}\label{eq:app3-ex3-P[q,1]=}
	\mathrm{P}[\obs; 1] \simeq \exp\left[ \sqrt{t}\; G\left(\frac{\obs}{\sqrt{t}}\right)\right],
	\quad t\to\infty,
\end{equation}
where
\begin{equation}\label{eq:app1-ex3-G(q)=}
	G(\eta) = \min_{\lambda} \left( - \eta\lambda + F(\lambda) \right)
\end{equation}
and
\begin{equation}\label{eq:app1-ex3-F(lambda)=}
	F(\lambda) = \frac{1}{\pi} \int_{-\infty}^{\infty} \dd k 
		\log\left[ 1 + \left(e^{\lambda}-1\right)e^{-k^2} \right]
\end{equation}
Expressions \eqref{eq:app1-ex3-G(q)=} and \eqref{eq:app1-ex3-F(lambda)=} give a parametric representation for $\mathrm{P}[\obs;1]$.

\end{appendix}

\bibliography{SC_v2.bib}

\end{document}